\documentclass[aps,twocolumn, secnumarabic,nobalancelastpage,amsmath,amssymb,nofootinbib,floatfix]{revtex4-1}

\usepackage{color}
\usepackage{graphics} 
\usepackage{graphicx} 
\usepackage{subfigure}
\usepackage{longtable} 
\usepackage{bm} 
\usepackage{epstopdf}
\usepackage[normalem]{ulem} 

\newcommand{\beq}{\begin{equation}}
\newcommand{\eeq}{\end{equation}}
\newcommand{\bqa}{\begin{eqnarray}}
\newcommand{\eqa}{\end{eqnarray}}

\newcommand{\cut}[1]{} 

\newcommand{\BQIC}{Berkeley Center for Quantum Information and Computation, Berkeley, California 94720 USA}
\newcommand{\DeptPhys}{Department of Physics, University of California, Berkeley, California 94720 USA}
\newcommand{\DeptChem}{Department of Chemistry, University of California, Berkeley, California 94720 USA}

\begin{document}

\title{Optimality of feedback control for qubit purification under inefficient measurement}

\author {Yuxiao Jiang$^{1,3} $}
\thanks{These authors contributed equally to this work}
\author{Xiyue Wang$^{1,2}$}
\thanks{These authors contributed equally to this work}
\author{Leigh Martin$^{1,2}$}
\author{K. Birgitta Whaley$^{1,3}$}
\email{corresponding author: whaley@berkeley.edu}

\affiliation{$^1$\BQIC}
\affiliation{$^2$\DeptPhys}
\affiliation{$^3$\DeptChem}

\date{\today}

\begin{abstract}
A quantum system may be purified, \textit{i.e.}, projected into a pure state, faster if one applies feedback operations during the measurement process. However existing results suggest that such an enhancement is only possible when the measurement efficiency exceeds 0.5, which is difficult to achieve experimentally. We address the task of finding the global optimal feedback control for purifying a single qubit in the presence of measurement inefficiency. We use the Bloch vector length, a more physical and practical quantity than purity, to assess the quality of the state, and employ a backward iteration algorithm to find the globally optimal strategy. Our results show that a speedup is available for quantum efficiencies well below 0.5, which opens the possibility of experimental implementation in existing systems. 
\end{abstract}

\maketitle

\section{Introduction}
\label{sec:introduction}
To implement quantum information processing, it is essential to have full knowledge of the quantum system at hand. When the state of a system is perfectly known, it is referred to as a pure state. However, under the influence of environmental noise or imperfect operations on the quantum state, a pure state will experience relaxation and decoherence, which in general can result in transformation to an unknown, or mixed state. While quantum error correction can be used to protect a computation or other specific operations on the qubit from decoherence, pure state initialization, \textit{i.e.}, generation of pure states from arbitrary mixed states is a far more basic task that underpins this, (e.g., in the production of clean ancillas), as well as countless other elementary tasks. 

A simple way to implement state purification is to perform a measurement. Although the elementary model of a measurement is a sudden interaction with a measurement device that gives rise to instantaneous wave function collapse, realistic measurements always take a finite amount of time and do not generally provide complete collapse. In the limit of a continuous measurement, information about the system state flows continuously from system to observer at a finite rate. 
Jacobs showed in 2003 that under continuous monitoring, when the measurement process is slow relative to the timescale of available unitary operations, 
feedback can increase the purification rate of qubits over what would be possible using measurement alone \cite{Jacobs2003}. The task of finding the optimal control protocol for qubit state purification under continuous measurements has since been studied in a number of works \cite{Jacobs2003,Wiseman2006reconsidering,Wiseman2008,Combes:2011wt}.
In discussions of such optimizations, it is important to understand the differences between locally and globally optimal control strategies.  A locally optimal strategy refers to a control strategy that optimizes the figure of merit at an instant in time.
In contrast, a globally optimal strategy provides the control protocol 
that optimizes the figure of merit over the finite time period between the allotted start and finish of the control task. 
The resulting instantaneous control protocol at any given time within this period will generally not be equal to the corresponding locally optimal control protocol. 

In Ref. \cite{Wiseman2006reconsidering}, Wiseman
and Ralph emphasized the 
distinction between two goals in qubit purification: i) the max purity goal, \textit{i.e.}, maximizing the purity at final time, and ii) the min time goal, \textit{i.e.}, minimizing the time required to achieve a given purity. Subsequent work 
established the globally optimal strategies for these two goals in the limit of perfect measurement efficiency, $\eta =1$ \cite{Wiseman2008}.
Li \textit{et al.} first discussed the global (over a finite time period) and local  (instantaneous) optimality of qubit purification in the presence of finite measurement inefficiency, $\eta < 1$ \cite{Li2013}. Using the verification theorem, they proved that the optimal strategy for the max purity goal with efficiency $\eta$ up to $1/2$ is to measure the state without applying feedback. 
This result is significant because routinely achieved quantum efficiencies are around $0.4$, and efficiencies well above $0.5$ would be required to demonstrate a significant purification enhancement. Many other imperfections like finite coherence time can be ameliorated by increasing the measurement rate, but efficiency is limited by challenging hardware constraints.

The choice of figure of merit for such a control problem strongly affects the optimal protocol, due to the stochastic nature of the system's evolution. Up to now, most work has used the state purity metric, defined as $\text{Tr[}\rho^{2}\text{]}$, to measure the quality of the state during the purification process. In practice however, as recognized in Ref.~\cite{Wiseman2008}, the length of the Bloch vector $r$ is arguably more relevant to many applications, 
primarily because it 
is linearly related to the fidelity between the instantaneous state and a target pure state, whereas the purity is nonlinear in this fidelity and not directly observable.
However an analytic solution for the globally optimal protocol under the Bloch vector length metric does not appear to be available except for the case of unit efficiency \cite{Wiseman2008}, which has impeded further investigation to the realistic situation of finite measurement inefficiency.

In this paper, we use Bloch vector length instead of purity as a metric to quantify state purification, and determine both the local and global optimality of control strategies for state purification in the presence of inefficient measurements. We develop a general method to numerically search for the globally optimal strategy and apply it to find a lookup table for the optimal control protocol. Our main result is that unlike the purity metric, the optimization according to the Bloch vector metric $r$ benefits from feedback control for any measurement efficiency $\eta\in [0,1]$.  To simplify the search process, we further make use of the stochastic version of Pontryagin's maximum principle (Appendix A) to restrict the control landscapes. The robustness and error estimates for our methods are discussed in Appendix B.  In the main text we take $\eta=0.3$ as an example to demonstrate the performance of the optimal protocol. Results for other values of $\eta$, both below and above 1/2, are provided in Appendix C. 
\section{Continuous Measurements and Feedback on a Qubit}
\label{sec:Setup}

We consider a qubit system subjected to weak, continuous measurement of operator X with measurement efficiency $\eta$ and strength $k$. The evolution of the system is described by stochastic master equation (SME):
\begin{equation}
\label{eq:SMEgeneral}
\begin{aligned}
d\rho &=  2k\mathcal{D}[X]\rho dt + \sqrt{2\eta k}\mathcal{H}[X]\rho dW.
\end{aligned}
\end{equation}
where $\mathcal{D}[A]\rho=A\rho A^{\dagger}-\frac{1}{2}(A^{\dagger}A\rho+\rho A^{\dagger}A)$ and $\mathcal{H}[X]=A\rho+\rho A^{\dagger}-\text{Tr}[(A+A^{\dagger})\rho]\rho$. $dW$ is a zero-mean, Gaussian random variable with variance $dt$. We assume the system evolves without the influence of relaxation and decoherence. The information that flows from system to observer is denoted by $dR=\sqrt{4k}\langle X\rangle dt+dW/\sqrt{\eta}$. The assumption that unitary operations may be implemented quickly relative to the measurement timescale implies that any feedback control can effectively be implemented instantaneously and with no delay. This means after each measurement time step, feedback applies a time-dependent unitary operation $U_{t}$ on the state such that $\rho(t+dt)=U_{t}\rho(t) U_{t}^{\dagger}$.

We now apply Eq.~\eqref{eq:SMEgeneral} to a qubit using the Bloch sphere representation, where any single-qubit density operator can be identified with a Bloch vector $\bm{r}=(x,y,z)$ in $\mathbb{R}^{3}$ using the relation $\rho=(\mathbb{I}+\bm{r}\cdot\bm{\sigma})/2$. 
Measurement perturbs the qubit state so that in the Bloch sphere representation, the dynamical evolution will appear as a random trajectory in $\mathbb{R}^{3}$. Without loss of generality, we consider here the purification effect of a weak measurement $X=\sigma_{z}/2$, starting from an initial state with zero $y$ component. We take further advantage of the symmetry of the Bloch sphere by assuming the unitary feedback operators to be rotations about the $y$ axis, so that the trajectory can be restricted to the $xz$ plane at all times. Under this simplification, each feedback protocol can be described with a single parameter $\theta(t)$, where at each time step the feedback control operator puts the state vector at some angle $\theta$ from the $+z$ axis. We can then parameterize each protocol by its function $u(r,t)$, where 
\begin{equation}
\label{eq:u_def}
u(r,t) = \text{cos}\theta(t) = \frac{z(t)}{r(t)},
\end{equation}
with $r$ the length of Bloch vector, which is our figure of merit. The actual unitary control operation is then of the form $U_{t}=\text{exp}\{-i\alpha(t)\sigma_{y}\}$, where $\alpha(t)$ is determined by a combination of the instantaneous state and $\theta$.  Note that the control value $u(r,t) = 1$ corresponds to no-feedback, and all other values including $u(r,t)=0$ correspond to active feedback.

Substituting the measurement operator $X=\frac{\sigma_{z}}{2}$ into Eq.~\eqref{eq:SMEgeneral}, we can isolate the $x$ and $z$ components of $d\rho$. Using  $r=\sqrt{x^{2}+z^{2}}$ and It\^{o}'s rule, we arrive at an equation of motion for 
$r$ \cite{Li2013}
\begin{equation}
\label{eq:SMEr}
\begin{aligned}
dr = k(r-\frac{\eta}{r})(u^{2}-1) dt + \sqrt{2k\eta}(1-r^{2})u dW,
\end{aligned}
\end{equation}
where we have suppressed the arguments of the variables $r(t)$ and $u(r,t)$. For notational convenience we shall continue to use the abbreviation whenever display of the arguments is not essential to the discussion. Under this setup, our optimal purification problem can be stated quite generally as minimizing some cost function $\mathcal{C}[u]$ given an initial state $r_0$, where $\mathcal{C}$ is a functional of the control function $u(r,t)$, $\mathbf{E}$ denotes the expectation value, and $T$ is the allowed duration of the purification protocol.  

For the ideal situation with perfect measurement efficiency, $\eta = 1$, measurement in a fixed (diagonal) basis without feedback has been identified as the globally optimal protocol for the ``min time'' goal \cite{Wiseman2006reconsidering}. Feedback that keeps the qubit unbiased with respect to the measurement basis \textit{i.e.}, ``always on" feedback, $u(r,t)=0$ for all $t$) is the globally optimal protocol for both the ``max purity'' and ``max Bloch vector length'' goals when $\eta = 1$ \cite{Wiseman2008,Combes:2011wt}. 

Less is known about the situation with inefficient measurements, $\eta < 1$. The no-feedback protocol (\textit{i.e.}, $u=1$) for the ``min time'' goal has been explicitly shown to be globally optimal for all values of $\eta$, with numerical evidence also showing that this global optimality is maintained in the presence of decoherence \cite{Li2013}. For the ``max purity'' goal with $\eta < 0.5$, measurement without feedback is known to be globally optimal (following an initial rotation to the measurement axis) \cite{Li2013}. However for $1/2 < \eta < 1$, the globally optimal protocol for the ``max purity'' goal is unknown. Furthermore, the optimal strategy for any non-unit value of $\eta$ is unknown for ``max Bloch vector length,'' arguably one of the most physically relevant goals.

As mentioned earlier, one of the motivations of this work is to overcome the impracticality of using state purity as a metric for purification, by using instead the Bloch vector length $r$ to measure the quality of the state. 
The cost function is then $\mathcal{C}[u]=\textbf{E[}1-r(T)\textbf{]}$, which is  
both measurable and linearly related to the state fidelity when the orientation of the Bloch vector is known. 

One may start by solving for the locally optimal protocol, \textit{i.e.}, the protocol that yields the greatest expected Bloch vector length increase $dr$ during a time interval $dt$. This can be done analytically; it amounts to maximizing the $dt$ term in Eq.~\eqref{eq:SMEr}, since the $dW$ contribution averages to zero. We thereby arrive at the locally optimal protocol
\begin{equation}
\label{eq:localOpt}
u_{lo,r}(r,t) = \begin{cases}
               0 & \text{if $r\leq r^{*}$} \\
                1 & \text{if $r> r^{*}$},
  				\end{cases}
\end{equation}
where $r^{*} = \sqrt{\eta}$ (see also Ref.~\cite{Combes:2011wt}). The same form was derived from a previous purification analysis of Eq.~\eqref{eq:SMEr} with the purity metric when the measurement inefficiency is greater than 0.5, \textit{i.e.}, $1/2<\eta\leq 1$, where $r^{*}_{p}=\sqrt{2-1/\eta}$  \cite{Li2013}.  However, with the purity metric, the locally (and global) optimal strategy was found to be $u_{lo,p}(r,t) = \pm 1$ when $\eta\leq 1/2$.

In contrast to the local optimality problem, the global optimality problem appears to be analytically intractable.  The difficulty arises in finding a general solution of Eq.~\eqref{eq:SMEr} for $r(T)$, given an arbitrary $u(r,t)$ and arbitrary initial conditions. Furthermore, the size of the search space is formidable, since $u(r,t)$ could be any function $f:[0,T]\times[0,1]\rightarrow[-1,1]$. However, the verification theorem \cite{Shabani2008} provides a way to numerically generate the globally optimal protocol while only exploring a small fraction of the full parameter space.  In addition, Pontryagin's maximum principle~\cite{yong1999stochastic} provides a useful necessary condition for the optimal control, namely that the control Hamiltonian must take an extreme value over all permissible controls. As shown in Appendix A, the stochastic form of the maximum principle can be used to show that for the measurement-based feedback approach to purification under the Bloch vector length cost function, the  optimal control is a discrete function $f:[0,T]\times[0,1]\rightarrow\{-1,0,1\}$, which significantly reduces the size of the search space. A further simplification comes from the observation that the control values $u=\pm1$ are physically equivalent. Therefore in the rest of this paper we need only consider the situations with $u=0$ and $u=1$ control values.

\section{Globally optimal control}

In this section we describe our numerical approach to finding the globally optimal strategy which maximizes $r$ at a final time $T$ in the presence of finite measurement inefficiency. For unit efficiency ($\eta=1$), the locally optimal strategy can be analytically solved (see Eq.~\eqref{eq:SMEr} above) and has been shown to be globally optimal by making use of the verification theorem \cite{Wiseman2008}. 
However, an analytic verification for inefficient measurements ($0< \eta <1$) is intractable, because in this regime neither $u=0$ nor $u=1$ is optimal, and integration of the equation of motion Eq.~\eqref{eq:SMEr}
is difficult to solve for non-constant $u$. 
Nevertheless, the principle of the verification theorem is sufficiently powerful to warrant development of numerical approaches. Here we present a general numerical method to search for the globally optimal strategy. In the following we first provide a general description of our approach and then apply it to the purification problem with arbitrary measurement efficiency.

We use a backwards propagation algorithm to iteratively minimize the cost function $C[u]$ to obtain the globally optimal protocol and the associated cost $C_{g}$.
We discretize the continuous variable $t$ into $M$ time steps, which are labelled by the upper index $j$ ($t^{j}=j\Delta t + t^0$, $\Delta t = T/M$). Suppose that $u(r,t)$ is piecewise constant over each time step $\Delta t$, so that we may use the notation $u^{j}=u(r,t^j)$. Under this constraint, the globally optimized cost function $C_g$ over a set of functions $\{u^{j}(r)\}_{j=1,..,M}$ is given by
\begin{eqnarray} \label{eq:GlobalCost}
C_{g}&=&\min_{u^{1},...,u^{M}}\int C(r^{M})P(r^{M}|r_{0},u^{1},...u^{M})d{\bf r}^{M}\nonumber\\
&=&\min_{u^{1},...,u^{M}}\iint
C(r^{M})P(r^{M}|r^{M-1},u^{M})\times \nonumber\\
&&\qquad\qquad\qquad P(r^{M-1}|r_{0},u^{1},...u^{M-1})d{\bf r}^{M-1}dr^{M}\nonumber\\
&=&\min_{u^{1},...,u^{M-1}}\int \tilde{C}^{M-1}(r^{M-1}) \times \\ \nonumber
&&\qquad\qquad\qquad
P(r^{M-1}|r_{0},u^1,...,u^{M-1})d{\bf r}^{M-1},
\end{eqnarray}
where we have defined $d\mathbf{r}^n \equiv \prod_{i=1}^n dr^i$. We have also defined the \textit{cost-to-go} recursively as
\begin{equation}
\label{eq:CostToGo}
\tilde{C}^{j-1}(r^{j-1}) \equiv\min_{u^{j}}\int \tilde{C}^{j}(r^{j})P(r^{j}|r^{j-1},u^j)dr^{j}.
\end{equation}
and take the base case as $\tilde{C}^{M}(r^M) \equiv C(r^M) = 1-r^M$. 
Here $C(r^{j})$ is the value of the cost function at time step $j$ and $P(r^{j}|r_{0},u^1,...,u^{j})$ is the probability distribution of $r$ at time step $t^j$, conditioned on some initial value $r_0$ and on the controls. We use the notation $d{\bf r}^k \equiv dr^1...dr^{k-1}dr^k$ to simplify the form of the equations.

The form of Eq.~\eqref{eq:GlobalCost} clearly illustrates the nature of the globally optimal control strategy, as a sequence of instantaneous control protocols $\{ u^{j} \}$ that applied sequentially in time will generate the optimal value of the cost function, $C_g$.  The $u^{j}$ control fields are determined as a whole and are not in general equal to the locally optimal control fields at the corresponding times $t^j$.

The  numerical optimization of $C_g$ proceeds as follows.  Starting with $j=M$, we first perform 
the minimization in the last line.  For this special case, the minimization is actually equivalent to finding the locally optimal protocol $u^M$, since $\tilde{C}^M(r^M)$ is the cost-to-go evaluated over only a single time step.
Since the cost function at the final time is known ($C(r^{M})=1-r^{M}$), using this together with the time evolving conditional probability distribution $P(r^{j}|r_{0},u^1,...,u^j)$, we are then able to calculate the overall cost function in an iterative procedure. We emphasize that although we write $r$ here as a continuous variable, in order to implement the algorithm numerically we replace the integral of $r$ by a Riemann sum with interval $\Delta r$, resulting in a discrete representation of the configuration and control parameter spaces $r^j$ and $u^{j}$ respectively.

Provided that $u$ is constant over each individual time interval $\Delta t$, the probability distribution function $P(r^{j}|r^{j-1},u^{j})$ can be calculated analytically from Eq.~\eqref{eq:SMEr} using the Fokker-Planck equation. As we claimed before, due to Pontryagin's maximum principle, we need only consider the cases $u=0$ or $1$. 
Within each time step we then optimize the cost function by evaluating the probability distribution of $r$ under the two different choices for $u$.

When $u=0$, the evolution is deterministic and we find
\begin{equation}
\label{r0}
r^{j}=\sqrt{\eta-(\eta-(r^{j-1})^{2})e^{-2k\Delta t}}.
\end{equation}
Since $r$ is discrete in the numerical implementation of our algorithm, we use a finite width $\delta$ function (implemented by a narrow Gaussian) to emulate the distribution $P(r^{j}|r^{j-1})$ in presence of feedback control ($u=0$):
\begin{equation}
\label{p1}
P(r^j|r^{j-1},0)=\frac{1}{\mathcal{N}}e^{-\frac{(r-r^{j})^{2}}{2\sigma^{2}}},
\end{equation}
where $\mathcal{N}$ is the usual Gaussian normalization factor. Thus for $u=0$, we propagate a $\delta$ function centered at $r^{j}$ with finite but small width $\sigma=\Delta r$.

The $u=1$ case is equivalent to measuring along the direction of the quantum state (which we assume is aligned along the $+z$ axis), without any feedback controls. The differential equation for the projection of the quantum state along the $z$ axis, $z(t)$, is solved in this case by \cite{Jacobs2004optimal}
\begin{equation}
\label{zu=1}
z(t)=  \frac{\sinh(\sqrt{2k\eta}W)+z_{0}\cosh(\sqrt{2k\eta}W)}{\cosh(\sqrt{2k\eta}W)+z_{0}\sinh(\sqrt{2k\eta}W)},
\end{equation}
where W is a random variable with distribution function 
\begin{align}
\label{p2}
P&(W|z_{0})   \\ \nonumber
&=\frac{1}{\sqrt{2\pi t}}e^{-\frac{W^{2}}{2t}-\eta kt}[\cosh(\sqrt{2k\eta}W)+\sinh(\sqrt{2k\eta}W)z_{0}].
\end{align}
Changing variables from $W$ to $z$ in Eq.~\eqref{p2}, we then obtain the probability distribution function of $z$, which we denote by $F(z,t|z_{0})$. Since the state remains aligned along the $z$ axis, \textit{i.e.}, $r=|z|$, we arrive at the following probability distribution function for $r$:
 \begin{align}
\label{eq:ru=1}
P(&r^j|r^{j-1},1) \\ \nonumber
&= \begin{cases}
               F(r^j,\Delta t|r^{j-1}) & \text{if $r^j=0$} \\
               F(r^j,\Delta t|r^{j-1})+F(-r^j,\Delta t|r^{j-1}) & \text{if $r^j>0$},
  				\end{cases}
\end{align}
where the $r^j=0$ case appears in order to avoid double-counting the probability of being at the origin, and the $F(-r^j,\Delta t|r^{j-1})$ term in the second line derives from the contributions with $z<0$. 
We summarize the searching process for $u^j$ as follows:
\begin{itemize}
  \item Calculate $P(r^{j}|r^{j-1},u^{j})$ from Eqs.~\eqref{p1}-\eqref{eq:ru=1} for all $r^{j-1}$ between 0 and 1, and both possible control values $u^{j}=0,1$.
  \item Compute the cost-to-go for each control value $u^j$ at this time step as $\tilde{C}_{u^j}^{j-1}(r^{j-1})= \displaystyle{\sum_{r^{j}}} \tilde{C}^{j}(r^{j})P(r^{j}|r^{j-1}, u^j)$.
  \item Take $\tilde{C}^{j-1}(r^{j-1})=\min\limits_{u}\tilde{C}^{j-1}_{u^j}(r^{j-1})$ \textit{i.e.}, Eq.~\eqref{eq:CostToGo} and begin the next round.
\end{itemize}
This procedure allows us to propagate the searching process backwards in time, with the final outcome being an explicit numerical solution for the optimal control strategy $u(r,t)$, represented as a discrete, two-parameter lookup table. By construction, this lookup table will provide the optimal control strategy for all possible values of $r_0 = r(t_0)$.  

\section{Results}
\begin{figure}[ht]
\centering
\includegraphics[width=0.5\textwidth]{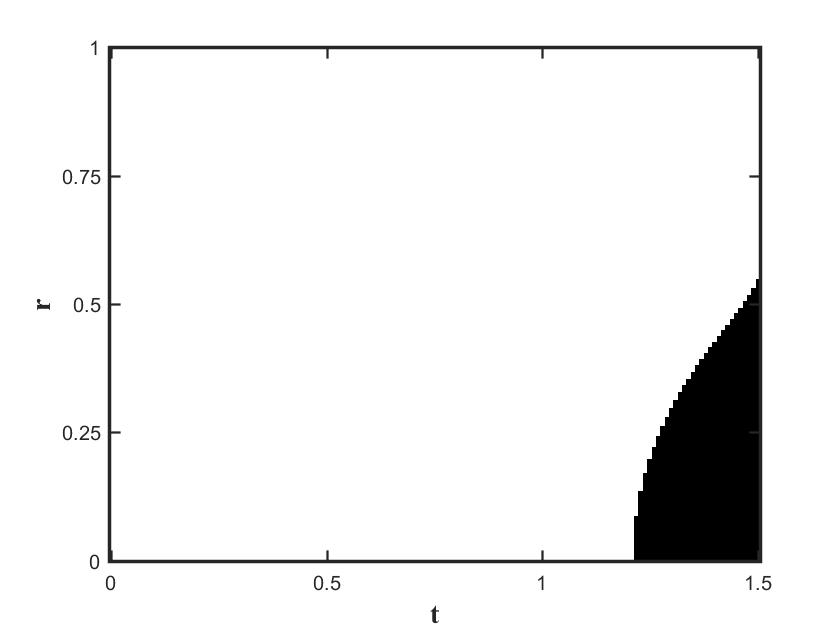}
\caption{Optimal control lookup table for measurement efficiency $\eta=0.3$. The control variable $u(r,t)$ takes  the value $0$ (apply feedback) in the black region and the value $1$ (no-feedback applied) in the white region. The feedback control is turned on for large $t$ and small $r$. For any allowed total purification time $t=T<1.5$, the time range [1.5 - T,1.5] provides a complete description of the control protocol. During the purification process, the state will follow a stochastic trajectory with the optimal control operation fully determined by the values of $(r,t)$.}
\label{controleta3}
\end{figure}

We choose an intermediate value of efficiency, $\eta=0.3$, as an example to illustrate and analyze our results. This is somewhat below the realistic values of measurement efficiency for superconducting qubit technologies today~\cite{hacohen2016noncommuting}.  Results for other efficiencies may be found in Appendix C. In Fig. \ref{controleta3}, we plot the optimal value of $u$ as a function of $r$ and $t$, for a total purification time $T=1.5$. This fully parametrizes our control strategy over the time interval $[0, 1.5]$ and provides a lookup table that can be used to generate the optimal control protocol for purification starting from any Bloch vector value $r(t_0)$ and evolving for a time $T = 1.5 - t_0 \leq 1.5$. The control values $u(r,t)$ are simply read off the table as time proceeds from $t=t_0$ to $t=t_0+T = 1.5$. 
For longer purification durations, the backwards recursion procedure would be made starting at a larger $T$ value.  Fig. \ref{controleta3} suggests that for all $T > 1.5$, the optimal strategy  will be to omit feedback at early times, meaning that the lookup table takes a value of $u=1$ for all times earlier than $t=T-1.5$.  Verifying this is however challenging, due to the numerical evaluation of the cost-to-go function, Eq.~\eqref{eq:CostToGo}.

We find a smooth boundary between regions in which one should apply feedback to align the state along the $x$ axis ($u=0$, black region), and regions where one merely measures parallel to the state ($u=1$, white region). The parameter region where feedback is required is centered around $r\rightarrow 0$, $t\rightarrow T$. This implies that to obtain the best possible value of the Bloch vector length $r$ at time $T$, feedback should be turned on only when we still have a small value of $r$, even if the system is already close to the final time. This derives from the fact that as $t$ approaches $T$, the globally optimal strategy in region $[t,T]$
approaches a locally optimal strategy. At $t\approx T$, the boundary between ``feedback'' and ``non-feedback'' regions is located exactly at $r = r^{*} = \sqrt{\mu}$ in the equation for the locally optimal strategy, Eq.~\eqref{eq:localOpt} and global and local optimality in $[t,T]$ thus become asymptotically equivalent to each other as this special point is approached. 

This situation differs from the optimal protocol when using purity as the figure of merit, where for $\eta<0.5$ the globally optimal strategy is $u=1$, \textit{i.e.}, measuring the state without any feedback~\cite{Li2013}. In contrast, Fig. \ref{controleta3} indicates that feedback does indeed benefit the purification rates under low efficiency measurements when we instead use the Bloch vector length $r$ to measure the quality of the state. Furthermore, when the available purification time $T_a$ is restricted to be significantly less than the specified time for the globally optimal protocol (here $T=1.5$), e.g., for $T_a \leq 0.3$, feedback will be an essential component of the optimal procedure when starting from the mixed state $r(t_0)=0$. However, the final value of $r(T_a)$ will thereby be reduced. When a longer duration can be used, the optimal protocol will only require feedback if and when the Bloch vector length $r(t)$ crosses the boundary to the $u=0$ (black) region at small $r$ values as the final time is approached. 
 
\begin{figure}[ht]
\centering
\includegraphics[width=0.5\textwidth]{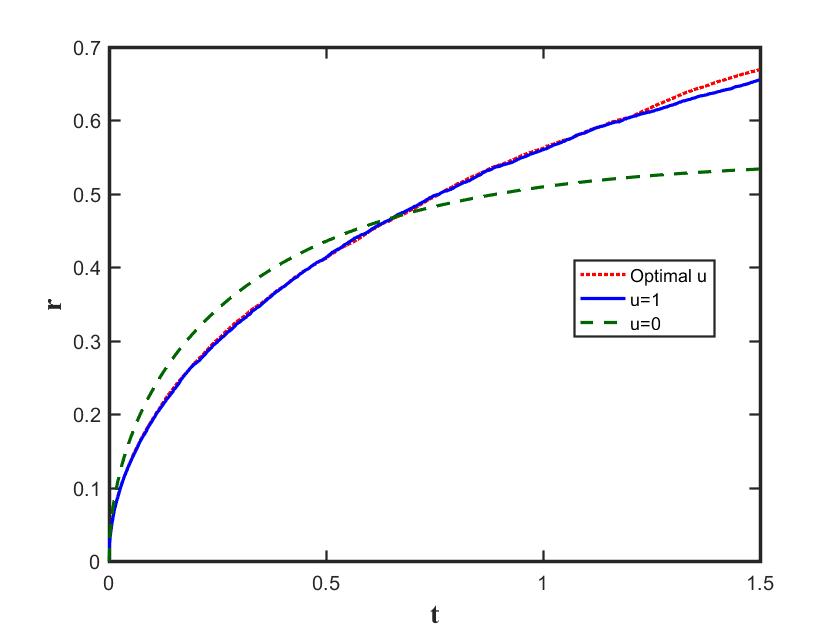}
\caption{Average value of the Bloch vector length $\langle r \rangle$ vs. time for the globally optimal strategy (red dotted line), for the $u=1$ protocol (no-feedback - solid blue line) and for the $u=0$ protocol (feedback always on - dashed green line)
for purification with measurement efficiency $\eta=0.3$ and measurement strength $k=1$, over a total purification time $T=1.5$. The initial condition here is the totally mixed state, $r_{0}=0$. Evolution under first two strategies is stochastic, according to Eq.~\eqref{eq:SMEr}, and we take the average over 10,000 trajectories. Evolution under the $u=0$ protocol is deterministic and is given by Eq.~\eqref{r0}.}
\label{r&t}
\end{figure}

In order to illustrate the benefits of the new control protocol, we simulate trajectories of $r$ using Monte Carlo method and evaluate $\langle r(t)\rangle$, where $\langle \cdot\rangle$ is the average over many trajectories. We compare the results achieved under three different strategies: $u=0$ (feedback always on), $u=1$ (measurement without feedback) and the globally optimal strategy. (Recall that the $u=0$ protocol is known to be globally optimal for the special case of perfect measurement efficiency, $\eta=1$ \cite{Wiseman2008}). The advantage of the globally optimal control strategy is clearly evident in Fig. \ref{r&t}, where we plot the results for a total purification time $T=1.5$. We see that until $t\approx 1.2$, the average value $\langle r(t)\rangle$ for the globally optimal strategy (red dotted line) is the same as the no feedback strategy (blue solid line). However, when the system reaches the critical time point $t=t^{*}\approx 1.2$ in the control lookup table of Fig. \ref{controleta3}, the optimal control strategy will switch on feedback for those trajectories having small $r(t)$ values. This causes the globally optimal outcomes to increase faster than those for the no feedback $u=1$ strategy, resulting in a larger value of the Bloch vector length for the globally optimal strategy at the final time, $r(T)$.  

It is important to note that while Fig. \ref{controleta3} specifies the globally optimal strategy for any final time $T < 1.5$, the curves in Fig. \ref{r&t} are specific to the final time value $T=1.5$. For smaller values of $T$, feedback would be activated at an earlier time and the difference between protocols would be larger.
%
\begin{table*}
\centering
\label{tab:cg&cmc}
\begin{tabular}{ccccccccccc}
\toprule
$\eta$ & 0.1 & 0.2 & 0.3 & 0.4 & 0.5 & 0.6 & 0.7 & 0.8 & 0.9 & 1\\
$C_{\text{MC}}$ & 0.5763 &  0.4290 &    0.3310  & 0.2601 & 0.2048  &  0.1611 & 0.1263  &  0.0967  &  0.0681  &  0.0255
\\
$\delta C_{MC}(\%)$  & 0.56 &   0.83 & 1.05  &  1.23  &  1.38 &   1.49 & 1.53   & 1.46 & 0.82  &  0.84\\
$\Delta(\%)$ & 0.19 &   0.29 &  0.20 &   0.21 & 0.05 &  0.07 &  0.07 & 0.03  & 0.01  & 0.03\\
\end{tabular}
\caption{Comparison of $C_{\text{MC}}$, the Monte Carlo averaged cost function evaluated on an ensemble of 10,000 trajectories evolving under Eq.~\eqref{eq:SMEr} with the globally optimal control protocol in which the numerically obtained value of the cost function $C_{g}$ is obtained by iterative minimization. 
$\delta C_{MC}$ is the standard error of the mean of $C_{MC}$ and $\Delta=|C_{g}-C_{\text{MC}}|/C_{\text{MC}}$ is the normalized difference between $C_{g}$ and $C_{\text{MC}}$.
The results are given for different values of the measurement efficiency $\eta$ between $0.1$ and $1$.  The 
quantum trajectories were run with measurement strength $k=1$, and initial condition equal to the totally mixed state, $r_{0}=0$.}
\end{table*}

We can validate our globally optimal control protocol by undertaking Monte Carlo trajectory simulations of the 
global cost function under the numerically obtained protocol $u(r,t)$. We denote the cost function value generated by Monte Carlo as $C_{\text{MC}}=1-\langle r(T)\rangle$, where $ \langle \cdot \rangle$ denotes the average over Monte Carlo trajectories. When a large number of trajectories are used to compute $C_{\text{MC}}$, the value should approach the value $C_{g}$ obtained from the backwards propagation algorithm. 
We confirm this relation in Table I, which shows a comparison between the values of $C_{\text{MC}}$ and $C_{g}$ for $\eta$ values ranging from $0.1$ to $1$. We simulate ten thousand trajectories under the optimal control strategy 
to obtain $C_{MC}$ and the associated uncertainty $\delta C_{MC}$, where the latter is estimated by the standard error of the mean.  Table I shows that the normalized difference 
$\Delta=|C_{g}-C_{\text{MC}}|/C_{\text{MC}}$ is consistently smaller than the uncertainty 
$\delta C_{MC}$, verifying the reliability of our results. 

\begin{figure}[ht]
\centering
\includegraphics[width=0.5\textwidth]{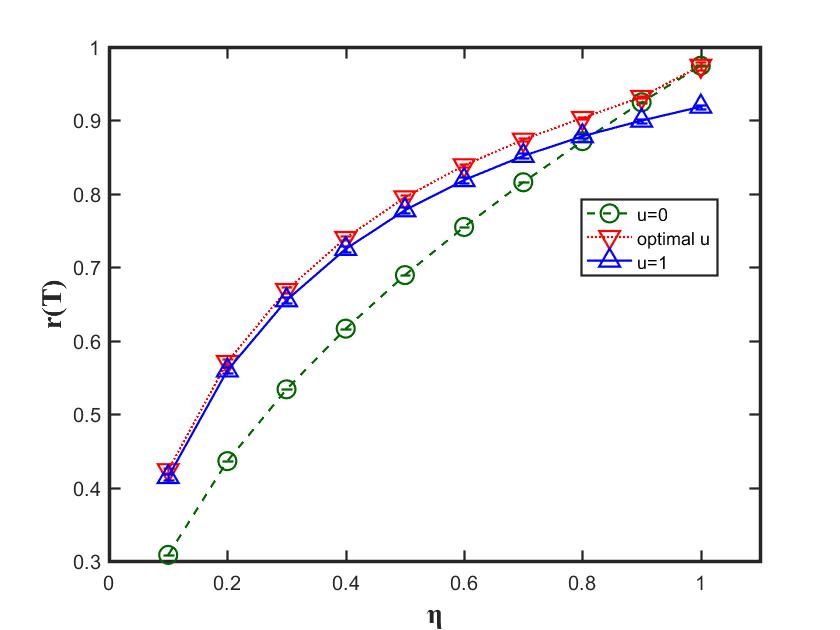}
\caption{Average value of the Bloch vector length $\langle r(T) \rangle$  
at the final time $T=1.5$ for the globally optimal strategy (red triangles), the $u(r,t)=0$ protocol (always on feedback, green circles) and the $u(r,t) =1$ protocol (no feedback, blue triangles), with measurement efficiencies $\eta$ varying from $0.1$ to $1$. All calculations were carried out with measurement strength $k=1$, and initial condition equal to the totally mixed state, $r_{0}=0$. Evolution under first two strategies is stochastic, according to Eq.~\eqref{eq:SMEr} and we take the average over 10,000 trajectories. Evolution under $u(r,t)=0$ protocol is deterministic and is given by Eq.~\eqref{r0}.}
\label{r&eta}
\end{figure}

To further demonstrate the benefits of the globally optimal protocol, we use the Monte Carlo trajectory simulation to calculate the final average Bloch vector value $\langle r(T)\rangle$ under different values of measurement efficiency for the three different control protocols, $u=0$, $u=1$, and the globally optimal protocol. A comparison of the three resultsover the  range from $\eta =0.1$ to $\eta =1$ for purification time $T=1.5$ is shown in Fig. \ref{r&eta}. We see that for finite but not perfect measurement efficiency, the globally optimal protocol always improves over both $u=0$ and $u=1$ control protocols. Furthermore, it approaches the no-feedback $u=1$ protocol as $\eta \rightarrow 0$ and approaches the always on $u=0$ feedback protocol as $\eta \rightarrow 1$.  Thus
at low $\eta$ values, the performance of the no-feedback protocol ($u=1$) is close to globally optimal, while the always on protocol ($u=0$) yields significantly lower values of $r(T)$. 
Indeed, from Eq.~\eqref{r0}, it is straightforward to show that in the long time limit $r\rightarrow \sqrt{\eta}$, independent of the initial value of $r$ (see also Ref.~\cite{Combes:2011wt}).
In contrast, at high $\eta$ values, the no-feedback $u=1$ protocol becomes increasingly ineffective, while the always on $u=0$ protocol becomes increasingly effective, resulting in a crossover of the corresponding $\langle r(T) \rangle$ values at $\eta \simeq 0.8$.  As $\eta$ further approaches unity, the globally optimal and always on ($u=0$) feedback protocols become increasingly close in value and converge at $\eta = 1$, as expected from the known equality of  local and globally optimal protocols in the case of perfectly efficient measurements~\cite{Wiseman2008} 
(see also Eq.~\eqref{eq:localOpt} and Fig. \ref{global04to1} in Appendix C).

In principle, it would now be possible to construct a plot of final Bloch vector length $\langle r(T)\rangle$ as a function of purification duration $T$, which could allow evaluation of the speedup of the globally optimal protocol relative to both the no feedback $u=1$ protocol and the always-on feedback $u=0$ protocol for finite efficiencies $\eta$.  However, since the final time $T$ is finite, asymptotic speedup factors cannot be extracted from such numerical solutions.  Nevertheless, Fig. \ref{r&eta} indicates that for $\eta < 1$ the globally optimal strategy will show a speedup over  the no-feedback $u=1$ strategy, since a larger value of $\langle r(T)\rangle$ is reached for a given time $T$.  The figure similarly indicates that there will be a speedup relative to the always-on $u=0$ strategy.

\section{Conclusion}
\label{sec:Conclusion}

This work shows that under all finite measurement efficiency $\eta$, including routinely achieved experimental values, quantum feedback accelerates qubit purification relative to purification via measurement alone. Our results identify the globally optimal feedback strategy for the full range of quantum measurement efficiencies, 
and show that an observable purification speedup should be achievable with existing technology employing less than 50\% measurement efficiencies. 
The globally optimal strategy approaches the no-feedback strategy as $\eta \rightarrow 0$ and the always on strategy as $\eta \rightarrow 1$, but performs better than both of these for all $0 < \eta < 1$.  
Even for finite purification duration times $T$, the globally optimal strategy is seen to show speedup relative to both of the constant control protocols in achieving purification to a given Bloch vector length $\langle r(T) \rangle$ close to unity.

Since no measurement is truly instantaneous, these results apply to a wide range of physical systems, although the greatest gains would come in systems with slower intrinsic measurement rates and demonstrated higher quantum efficiencies, such as superconducting qubits.

The present study indicates several clear goals for further research. Firstly, for a given 
physical architecture, a complete analysis of all imperfections should be undertaken to 
ensure that the demonstrated advantage of feedback outweighs the cost of implementing this. 
In superconducting circuits, the primary imperfections after the non-unit measurement 
efficiency are likely to be presence of feedback delay and also unwanted interactions 
between the feedback operations and the readout protocol, such as non-Markovian effects in 
circuit QED readout\cite{gambetta2008quantum}. 
Such non-Markovian effects are a rich and active topic of fundamental research, and our 
current work motivates a significant and novel application of this work to these settings.

Secondly, the methods implemented here are applicable to many related problems in quantum control. 
Even larger gains are available for purifying systems larger than a single qubit
\cite{Combes2008rapid,combes2011maximum}, although to our knowledge no work has addressed the effect of measurement inefficiency in the multi-qubit context. Measurement-based feedback for entanglement generation has also proven a fruitful avenue of 
research\cite{stockton2004dicke,HPFPRA,martin2017optimal,zhang2018locally}, and open problems regarding optimality remain even in the case of two qubit Bell state generation\cite{martin2017optimal}. 
Systems substantially more complex than a single qubit may require the use of approximations or restrictions on the feedback protocol, which can impede proof of global optimality. However these challenges should not discourage the search for better feedback protocols, particularly in cases where there is a large margin to gain over the corresponding no-feedback protocols. 
The fact that the benefits of measurement-based feedback persist in realistic scenarios with inefficient measurement bodes well for future applications.

\begin{acknowledgments}
LM was supported by the National Science Foundation Graduate Fellowship Grant No. (1106400) and the Berkeley Fellowship for Graduate Study. 
KBW was supported by
Laboratory Directed Research and Development (LDRD)
funding from Lawrence Berkeley National Laboratory,
provided by the U.S. Department of Energy, Office of
Science under Contract No. DE-AC02-05CH11231. 
\end{acknowledgments}

%

\section*{Appendix A: Stochastic extension of Pontryagin's Maximum Principle}
In this section, we show that the optimal control function $u(r,t)$ must take the values $0$ or $\pm 1$. The original version of Pontryagin's Maximum Principle is for deterministic problems and thus can not be directly applied to our problem of purification by measurement-based feedback. Instead, we use the Stochastic Maximum Principle \cite{yong1999stochastic} to prove our claim. Consider the stochastic master equation for a variable $x(t)$,
\begin{equation}
\label{eq:SME_Pontryagin}
dx(t)=A(t,x(t),u(t))dt+B(t,x(t),u(t))dW,
\end{equation}
with cost functional
\begin{equation}
J(u)=\textbf{E}\left[ \int_{0}^{T}f(t,x(t),u(t))dt, +h(x(T)) \right],
\end{equation}
where $dW$ is a stochastic increment, $u(t)$ the control function, and the cost functional depends on the time dependent function $f(t,x(t),u(t))$ and a function $h(x(T)$ dependent only on the final time $T$.
The Stochastic Maximum Principle states that the optimal $u(t)$ must satisfy the condition
\begin{equation}
\label{condition}
\mathcal{H}(t,\overline{x}(t),\overline{u}(t))=\max_{u(t)}\mathcal{H}(t,\overline{x}(t),u(t))
\end{equation}
where $\{\overline{x}(t),\overline{u}(t)\}$ defines the optimal protocol and 
the function $\mathcal{H}$ is defined below.  Before giving the explicit form of $\mathcal{H}$, we first define two pairs of adjoint variables $\{P(t),Q(t),q(t),p(t)\}$. 
The variables $P(t), p(t)$ are solved for by the following set of  
terminal value stochastic differential equations \cite{yong1999stochastic}
\begin{equation}
\label{solve1}
\left\{
    \begin{array}{lll}
    dp(t)&=&-\{A_{x}(t,\overline{x},\overline{u})^\intercal p(t)
    +\displaystyle{\sum_{j=1}^{m}}B^{j}_{x}(t,\overline{x},\overline{u})^\intercal q_{j}(t)\\
    &&-f_{x}(t,\overline{x},\overline{u})\}dt+q(t)dW\\
    \\
    p(T)&=&-h_{x}(\overline{x}(T)),
    \end{array}
\right.
\end{equation}
\begin{equation}
\label{solve2}
\left\{
    \begin{array}{lll}
    dP(t)&=&-\{A_{x}(t,\overline{x},\overline{u})^\intercal P(t)+ P(t)A_{x}(t,\overline{x},\overline{u})\\
    &&+\displaystyle{\sum_{j=1}^{m}}B^{j}_{x}(t,\overline{x},\overline{u})^\intercal P(t)B^{j}_{x}(t,\overline{x},\overline{u})\\
    &&+\displaystyle{\sum_{j=1}^{m}}\{B^{j}_{x}(t,\overline{x},\overline{u})^\intercal Q_{j}(t)+Q_{j}(t)B^{j}_{x}(t,\overline{x},\overline{u}) \}\\
    &&+H_{xx}(t,\overline{x},\overline{u},p(t),q(t))\}dt+\displaystyle{\sum_{j=1}^{m}}Q_{j}(t)dW^{j}\\
    \\
    P(T)&=&-h_{xx}(\overline{x}(T)),
    \end{array}
\right.
\end{equation}
where for notational convenience we have omitted the explicit time dependence of $\{\overline{x}(t),\overline{u}(t)\}$. We shall similarly drop the explicit time dependence in other functions below. 

The Hamiltonian $H$ in Eq.~\eqref{solve2} is given by
\begin{equation}
H(t,x,u,p,q) \equiv \langle p,A(t,x,u)\rangle + \text{tr}[q^\intercal B(t,x,u)]-f(t,x,u).
\end{equation}
Given the 6-tuple $x, u, P, Q, p, q$, we then define the function $\mathcal{H}$ as
\begin{equation}
\begin{aligned}
&\mathcal{H}(t,x,u)\\
\equiv &\frac{1}{2}\text{tr}[B(t,x,u)^\intercal P(t)B(t,x,u)]+\langle p(t),A(t,x,u)\rangle-f(t,x,u)\\
&+\text{tr}[q(t)^\intercal B(t,x,u)]-\text{tr}[B(t,x,u)^\intercal P(t)B(t,\overline{x},\overline{u})].
\end{aligned}
\end{equation}
Since all we need here is to find the solutions to Eq.~\eqref{condition}, it is unnecessary to get the complete solutions of these adjoint equations. 
We first note that any feedback controlled evolution must be ``adapted" in the sense that the feedback terms cannot depend on the measurement noise $dW$.  This implies that the 
terms $Q(t)$ and $q(t)$ in Eqs. \eqref{solve1} - \eqref{solve2} must be equal to zero.

Then replacing $x$ by $r$, with $A, B$ given by the purification equation of motion for $r$ in Eq.~\eqref{eq:SMEr} with cost functions $f=0, h=1-r(T)$ corresponding to maximizing the length of the Bloch vector, we find  
\begin{equation}
\begin{aligned}
\mathcal{H}(t,\overline{r},u)=2kP(t)\eta&(1-\overline{r}^{2})^{2}(\frac{u^{2}}{2}-u\overline{u})\\
&+p(t)k(\overline{r}-\frac{\eta}{\overline{r}})
(u^{2}-1)
\end{aligned}
\end{equation}
where $u\in[-1,1]$. Since $\mathcal{H}$ is a continuous function of $u$, the maximum value of $\mathcal{H}$ with respect to variations in $u$ can be reached at the boundary points $u = \pm 1$ and at any point where the derivative with respect to the control is zero, \textit{i.e.}, when 
\begin{align*}
 \mathcal{H}_{u} & =2k\eta P(t)(1-\overline{r}^{2})^{2}(u-\overline{u})+2p(t)k(\overline{r}-\frac{\eta}{\overline{r}})u \\ \nonumber
 & = 0
 \end{align*}
From this it is evident that $\mathcal{H}_{u}$ is zero only when $\overline{u}=u=0$. Thus the optimal control values at each time $t$ can be drawn only from the discrete set $\overline{u}= \{0, \pm 1 \}$.

\section{Appendix B: Error Analysis}
\label{sec:verification}

In this section we analyze the magnitude of numerical errors and their effects on the final results. The numerical error of our protocol mainly comes from the discretization of $r$ and $t$. Since we already use analytical solutions of the underlying stochastic differential equation, error from discretization of $t$ should be negligible. However, the magnitude of $\Delta r$ affects the accuracy of our protocols due to the fact that we use a Riemann sum to approximate the integration of the cost function. The approximation leads to uncertainty in the cost function, which could affect the optimal control lookup tables. In this section, we show a method to obtain the uncertainty in the cost function and use it to calculate the error of our control protocol.

Recall that the cost-to-go at time step $j$ is given by
\begin{equation}
\tilde{C}^{j-1}(r^{j-1})= \displaystyle{\sum_{r^{j}}} \tilde{C}^{j}(r^{j})P(r^{j}|r^{j-1})
\end{equation}
where minimization over $u^j$ is taken to be implicit. In this formula, $P$ is exact and the error comes from cost-to-go $\tilde{C}$ and the approximation of the Riemann sum. We write the error in $\tilde{C}^{j-1}(r^{j-1})$ as
\begin{equation}
\delta \tilde{C}^{2}(r^{j-1})=\displaystyle{\sum_{r^{j}}}
\delta \tilde{C}^{2}(r^{j})P^{2}(r^{j})+\delta^{2}_{R}
\end{equation}
where the first term describes the propagation of error from the previous time step and second term is the error caused by approximation of the Riemann sum at the current time step. It should be emphasized that at the first step $[T-dt,T]$, $\tilde{C}$ is exact, therefore the error all comes from $\delta_{R}$.

\begin{figure}[ht]
\centering
\includegraphics[width=0.5\textwidth]{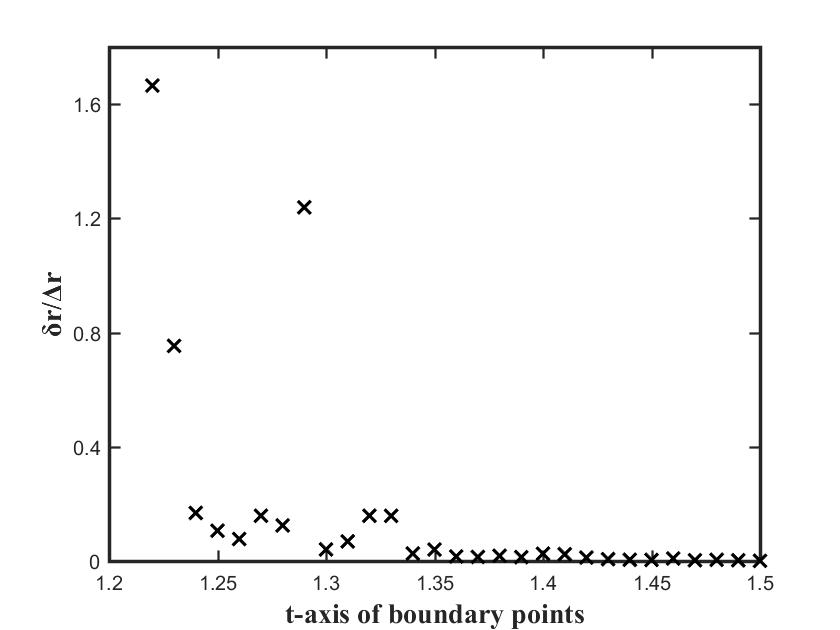}
\caption{Relative error in the length of the Bloch vector, $\delta r/\Delta r$, in the boundary points between feedback and no-feedback regions of the control lookup table, shown as a function of time $t$, for controlled evolution under measurement efficiency $\eta=0.3$. The parameters are the same as those in Fig.} \ref{controleta3}.
\label{er}
\end{figure}

We now explain the method for calculating $\delta_{R}$. When $u=1$, the error can be simply calculated using the standard formula for the error of a Riemann sum. When $u=0$, the distribution function $P$ will be a delta function and the standard formula fails. Instead, the error can be estimated via the fact that $\int C(r)\delta(r-r_{0})dr\approx C(r_{0})+\mathcal{O}(\sigma)$ where $\sigma$ is standard deviation of the approximate delta function used in practice. At each step we evaluate $\delta_{R}$ and thereby determine $\delta \tilde{C}_{u=0}(r,t)$ and $\delta \tilde{C}_{u=1}(r,t)$ completely.
Recall that at each step we determine the optimal $u$ by evaluating $\Delta \tilde{C}(r,t)=\tilde{C}_{u=0}(r,t)-\tilde{C}_{u=1}(r,t)$. Thus the error of $r$ in the position of the boundary points between the $u=0$ and $u=1$ regions is given by
\begin{equation}
\delta r= \pm \sqrt{\delta \tilde{C}_{u=0}^{2}+\delta \tilde{C}_{u=1}^{2}}\left(\frac{d\Delta \tilde{C}(r,t)}{dr}\right)^{-1}
\end{equation}
The result is shown in Fig. \ref{er}, where we have plotted the ratio of $\delta r$ to the instantaneous radial increment $\Delta r$, \textit{i.e.}, v $\delta r / \Delta r$ as a function of time for the  
boundary points in Fig \ref{controleta3}.
Choosing $\delta r/\Delta r$ helps us measure the relative error of boundary points compared to the increment $\Delta r$.
While the error of $r$ in most boundary points is less than the radial step increment
$dr$, there are some exceptional points with error larger than $\Delta r$. One may worry that such points might lead to divergence and make the whole protocol unstable. To check this, we choose several different values of $\Delta r$. We find the position of boundary points are identical, confirming the accuracy of our optimal control protocols.

The above error analysis is restricted to boundary regions. In other words, for all values of $\eta$, our method accurately provides a set of boundary points between feedback and non-feedback regions. However, due to numerical error, a complete error analysis of the control table within the feedback and non-feedback region is harder to predict, especially for high values of $\eta$.

\section{Appendix C: Optimal purification strategy }
\label{sec:LookupTable}

\begin{figure}[ht] 
\centering
\includegraphics[width=0.5\textwidth]{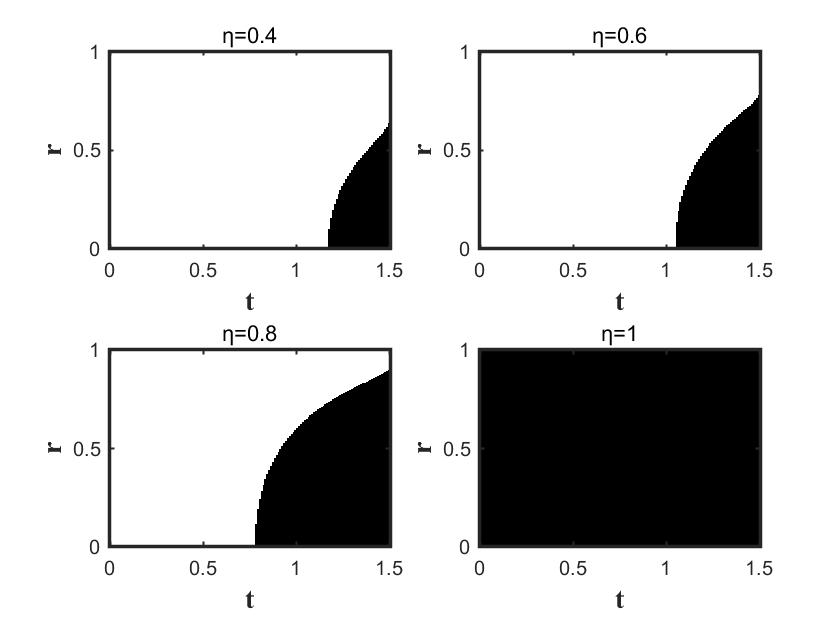}
\caption{Optimal control lookup tables for $u(r,t)$ at measurement efficiency values $\eta=0.4,0.6,0.8,1$. The black regions indicate feedback on, $u(r,t)=0$, and the white regions indicate {feedback off,} $u(r,t)=1$.}
\label{global04to1}
\end{figure}

We supplement the main text here with figures showing the globally optimal strategy lookup tables, \textit{i.e.}, $u(r,t)$, for different values of $\eta$. As is shown in Fig. \ref{global04to1}, for better measurement efficiency $\eta$, the feedback is activated earlier and covers a larger region. For perfect efficiency, $\eta=1$, the optimal control strategy found by our numerical optimization procedure is $u=0$ regardless of $r$ and $t$, which coincides with the analytical solution derived in previous work \cite{Wiseman2008}.

\clearpage
\end{document}